\documentclass[twocolumn,superscriptaddress,showpacs,prb]{revtex4}

\usepackage{graphicx}%
\usepackage{dcolumn}
\usepackage{amsmath}
\usepackage{bm}
\usepackage{color}

\begin{document}



\preprint{submitted to Physical Review Letter}
\title{
Bulk nature of layered perovskite iridates beyond the Mott scenario \\
: An approach from bulk sensitive photoemission study
}

\author{A. Yamasaki}
\author{S.~Tachibana}
\affiliation{Faculty of Science and Engineering, Konan University, Kobe 658-8501, Japan }

\author{H. Fujiwara}
\affiliation{Graduate School of Engineering Science, Osaka University, Toyonaka, Osaka 560-8531, Japan}

\author{ A.~Higashiya}
\affiliation{Faculty of Science and Engineering, Setsunan University, Neyagawa, Osaka 572-8508, Japan}
\affiliation{RIKEN SPring-8 Center, Sayo, Hyogo 679-5148, Japan}

\author{ A. Irizawa}
\affiliation{Institute of Scientific and Industrial Research, Osaka University, Ibaraki, Osaka 567-0047, Japan}

\author{ O. Kirilmaz}
\author{ F.~Pfaff}
\author{ P. Scheiderer}
\author{ J.~Gabel}
\author{M.~Sing}
\affiliation{Physikalisches Institut, Universit\"at W\"urzburg,  D-97074 W\"urzburg, Germany}

\author{T. Muro}
\affiliation{Japan Synchrotron Research Institute, Sayo, Hyogo 679-5198, Japan}

\author{M.~Yabashi}
\author{K.~Tamasaku}
\affiliation{RIKEN SPring-8 Center, Sayo, Hyogo 679-5148, Japan}

\author{H.~Sato}
\author{H.~Namatame}
\affiliation{Hiroshima Synchrotron Radiation Center, Hiroshima University, Higashi-Hiroshima, Hiroshima 739-0046, Japan}

\author{M.~Taniguchi}
\affiliation{Hiroshima Synchrotron Radiation Center, Hiroshima University, Higashi-Hiroshima, Hiroshima 739-0046, Japan}
\affiliation{Graduate School of Science, Hiroshima University, Higashi-Hiroshima, Hiroshima 739-8526, Japan}

\author{ A.~Hloskovskyy}
\affiliation{DESY Photon Science, Deutsches Elektronen-Synchrotron, D-22603 Hamburg, Germany}

\author{ H.~Yoshida}
\author{H.~Okabe}
\author{ M.~Isobe}
\affiliation{National Institute for Materials Science, Tsukuba, Ibaraki 305-0044, Japan}

\author{ J.~Akimitsu}
\affiliation{College of Science and Engineering, Aoyama Gakuin University, Sagamihara, Kanagawa 252-5258, Japan}

\author{W.~Drube}
\affiliation{DESY Photon Science, Deutsches Elektronen-Synchrotron, D-22603 Hamburg, Germany}

\author{ R. Claessen}
\affiliation{Physikalisches Institut, Universit\"at W\"urzburg,  D-97074 W\"urzburg, Germany}

\author{T. Ishikawa}
\affiliation{RIKEN SPring-8 Center, Sayo, Hyogo 679-5148, Japan}

\author{S. Imada}
\affiliation{College of Science and Engineering, Ritsumeikan University, Kusatsu, Shiga 525-8577, Japan}

\author{A. Sekiyama}
\affiliation{Graduate School of Engineering Science, Osaka University, Toyonaka, Osaka 560-8531, Japan}
\affiliation{RIKEN SPring-8 Center, Sayo, Hyogo 679-5148, Japan}

\author{ S. Suga}
\affiliation{Graduate School of Engineering Science, Osaka University, Toyonaka, Osaka 560-8531, Japan}
\affiliation{RIKEN SPring-8 Center, Sayo, Hyogo 679-5148, Japan}
\affiliation{Institute of Scientific and Industrial Research, Osaka University, Ibaraki, Osaka 567-0047, Japan}

\date{\today}

\begin{abstract}
We present genuine bulk Ir $5d$ $j_{\it eff}$ states of layered perovskite iridates obtained by hard-x-ray photoemission spectroscopy (HAXPES) with $s$- and $p$-polarized lights.
HAXPES spectra of Sr$_2$IrO$_4$ and Ba$_2$IrO$_4$ are well reproduced by the  quasi-particle densities of states calculated by the local density approximation with dynamical mean-field theory (LDA+DMFT).
It is demonstrated that the insulating nature of the iridates is triggered by antiferromagnetic correlation (Slater type) combined with electron correlation (Mott type).
The extremely-low-energy bulk-sensitive photoemission spectroscopy reveals ``bad metallic'' states in the paramagnetic phase of the iridates,
suggesting  strongly renormalized metallic states  above  the N\'eel temperature as predicted by the LDA+DMFT.

\end{abstract}


\maketitle

%
%


Mott physics, which explains the insulating nature in materials triggered by strong electron correlations,  has been vigorously studied for several decades from both experimental and theoretical points of view.~\cite{Imada98}
Owing to these efforts,  new mechanisms for the metal-insulator transition  (MIT) were proposed in some ``Mott'' systems, for instance, orbital switching for VO$_2$ and the essential role of  antiferromagnetic (AF) correlation effects for La$_2$CuO$_4$.~\cite{Haverkort05,Comanac08}

Layered perovskite iridates $A_2$IrO$_4$ ($A$=Sr, Ba) have an insulating state, the origin of which was mysterious in the early stage of the research.~\cite{Cava94}
Detailed spectroscopic and theoretical studies recently demonstrated Sr$_2$IrO$_4$ to be a spin-orbit(SO)-driven Mott insulator.~\cite{Kim08}
According to that scenario, the electronic structure near the Fermi level ($E_F$)  is characterized by Kramers-doublet  Ir 5$d$ $j_{\it eff}$=1/2 states, which are  eigenstates of the SO-hamiltonian within the $t_{2g}$ subspace.
For the iridate with five 5$d$ electrons, the lower $j_{\it eff}$=3/2 bands are fully occupied and the higher half-filled $j_{\it eff}$=1/2 band split  into the upper and lower Hubbard bands due to the relatively weak on-site Coulomb interaction ($U$).
This simple picture has been widely accepted since the results of  experiments, such as  optical conductivity spectra, seemed  to be well explained.~\cite{Kim08,Moon09}
In addition, the results of angle-resolved ultraviolet photoemission spectroscopy were well reproduced by the local density approximation (LDA)+$U$+SO band structure calculations.~\cite{Kim08}
For $5d$ transition-metal  systems, however, the nature of  atom-band duality, which is a characteristic feature of heavy elements, makes the interpretation of  optical conductivity spectra complicated.~\cite{BHKim12}
In addition, we point out here  that the photoemission spectra excited by several tens-eV photons are strongly affected by the surface electronic structures modified by stronger electron correlation effect on the surface than in the bulk even for  quasi two-dimensional compounds.~\cite{Sekiyama00,Sekiyama04}

Recently, the LDA study with dynamical mean-field theory  (LDA+DMFT)  pointed out that both  Sr$_2$IrO$_4$ and Ba$_2$IrO$_4$ should be classified as Slater insulators, in which the  AF correlation mainly contributed to  realize the insulating states in these materials rather than the electron correlation.~\cite{Arita12,Slater51}
Then, the time-resolved optical experiment  revealed in Sr$_2$IrO$_4$ two different characteristic behaviors, each of which could be  explained by Slater or Mott physics.~\cite{Hsieh12}

In this paper, we focus on the  layered perovskite iridates Sr$_2$IrO$_4$ and Ba$_2$IrO$_4$ which have comparable N\'eel temperature $T_N$ ($\simeq$ 250~K and 240~K) but different crystal symmetries ($I4_1$/{\it acd}  and $I4$/{\it mmm} ).~\cite{Crawford94,Okabe11}
It is an intriguing problem whether both iridates have a common insulating nature or not.
Detailed information on their  bulk electronic structures facilitates discussions about not only  Mott and Slater physics but also  a possibility of superconductivity, attracting a wide general interest.~\cite{Wang11,Watanabe13}
In order to provide much deeper insight into the nature of the SO-induced insulating states,
we have carried out  soft-x-ray and hard-x-ray photoemission spectroscopy (SXPES and HAXPES) in  the photon energy ($h\nu$) ranges of 400-1200~eV and 6-8~keV, respectively.
HAXPES reveals  genuine bulk $j_{\it eff}$ states owing to the  large  photoionization cross section ($\sigma$) of Ir $5d$ states~\cite{Lindau85} and  the long inelastic mean-free path ($\lambda_{\rm mp}$) of detected photoelectrons.~\cite{Yamasaki0710,Tanuma94}
These findings are compared  to the results of the LDA+DMFT calculations.~\cite{Arita12}
Furthermore, the extremely-low-energy photoemission spectroscopy (ELEPES) has been performed to reveal the bulk electronic states in the vicinity of  $E_F$.
ELEPES  demonstrates  that both Sr$_2$IrO$_4$ and Ba$_2$IrO$_4$ have ``bad metallic'' states above $T_N$ as predicted by the calculations.


HAXPES was performed at the long-undulator beamline BL19LXU in SPring-8~\cite{Yabashi_01} with the MB Scientific A-1~HE spectrometer for Sr$_2$IrO$_4$ and at the undulator beamline P09 in Petra III at DESY~\cite{Strempfer13} 
with the SPECS PHOIBOS 225 HV spectrometer for Ba$_2$IrO$_4$.
SXPES  was carried out at the twin-helical undulator 
beamline BL25SU in SPring-8 using the  GAMMADATA-SCIENTA SES-200 spectrometer.~\cite{RSI25}
The undulators in the SX and HAX beamlines produce  circularly and linearly polarized lights, respectively.
ELEPES experiments were performed with the MB Scientific T-1 xenon discharge lamp~\cite{Souma07,Suga10} and  the A-1 spectrometer at Konan University.
Inverse photoemission spectroscopy (IPES) was performed in HiSOR at Hiroshima University.~\cite{Sato98}
For all the measurements, polycrystalline samples  were used.~\cite{Okabe11}
Clean surfaces were obtained by fracturing 
samples {\it in situ} in UHV ($\leq$2$\times10^{-7}$ Pa for HAXPES, $\leq$3$\times10^{-8}$ Pa for SXPES, XAS, ELEPES, and IPES).
The total energy resolution ($\Delta E$) of each experiment was estimated by the Fermi edge of gold.

\begin{figure}
\includegraphics[width=7.0cm,clip]{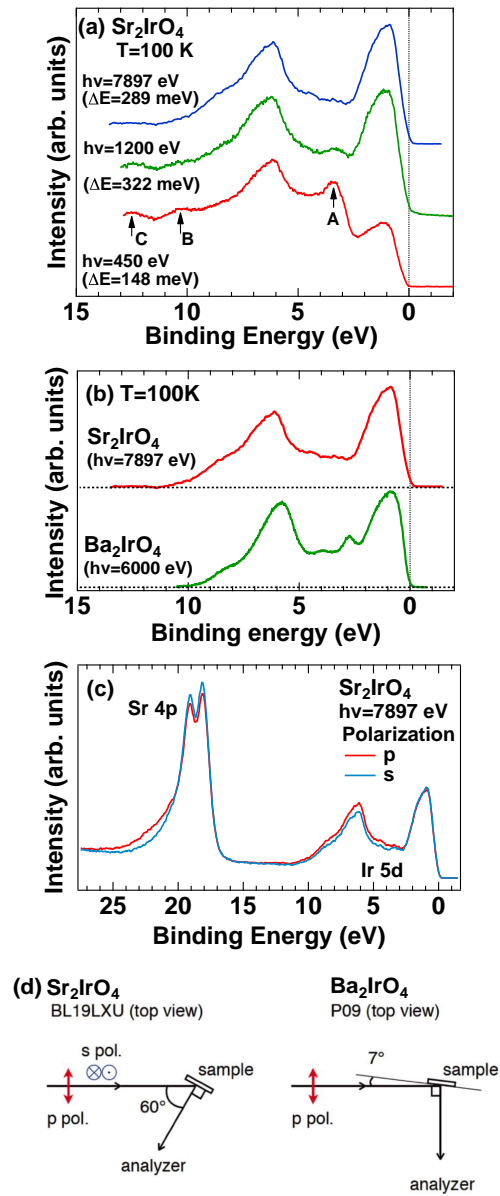}
\caption{
(Color online) 
(a)Valence-band SXPES and HAXPES spectra of Sr$_2$IrO$_4$.
(b)Valence-band HAXPES spectra of Sr$_2$IrO$_4$ and Ba$_2$IrO$_4$ after subtracting  Shirley-type backgrounds.
Both spectra were obtained for  $p$-polarized lights.
(c)Polarization-dependent HAXPES spectra of Sr$_2$IrO$_4$.
(d)Schematic illustrations of experimental setups for HAXPES.
}
\label{Fig_1}
\end{figure}
\begin{figure}
\includegraphics[width=7.0cm,clip]{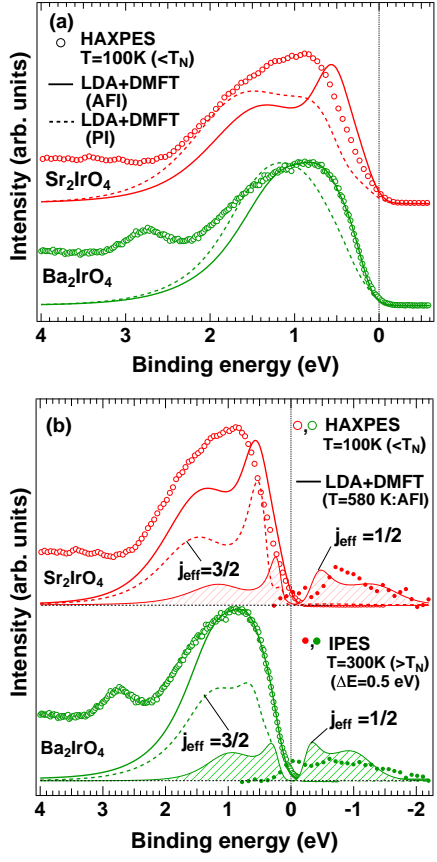}
\caption{
(Color online) 
Valence-band HAXPES spectra  in $j_{\it eff}$ states of Sr$_2$IrO$_4$ and Ba$_2$IrO$_4$.
(a)Calculated spectra for both  AFI (solid line) and  PI (dashed line) phases are  shown, which are broadened by Gaussian and Lorentzian functions representing the experimental energy-resolution and lifetime  effects.
Finite temperature effect around $E_F$ is also considered.
(b)IPES spectra measured for the kinetic energy of incident electrons, $E_K$=50~eV at $T$=300~K ({\it above} $T_N$) are added (indicated by solid circles). 
Calculated spectra of both  $j_{\it eff}$=3/2 (dashed line) and 1/2 (solid line with hatches) states for the AFI phase are also shown separately. 
All the experimental spectra are normalized by the area under the curves after subtracting  Shirley-type backgrounds(see  Fig.~\ref{Fig_1}(b)).
}
\label{Fig_2}
\end{figure}


\begin{figure}
\includegraphics[width=6cm,clip]{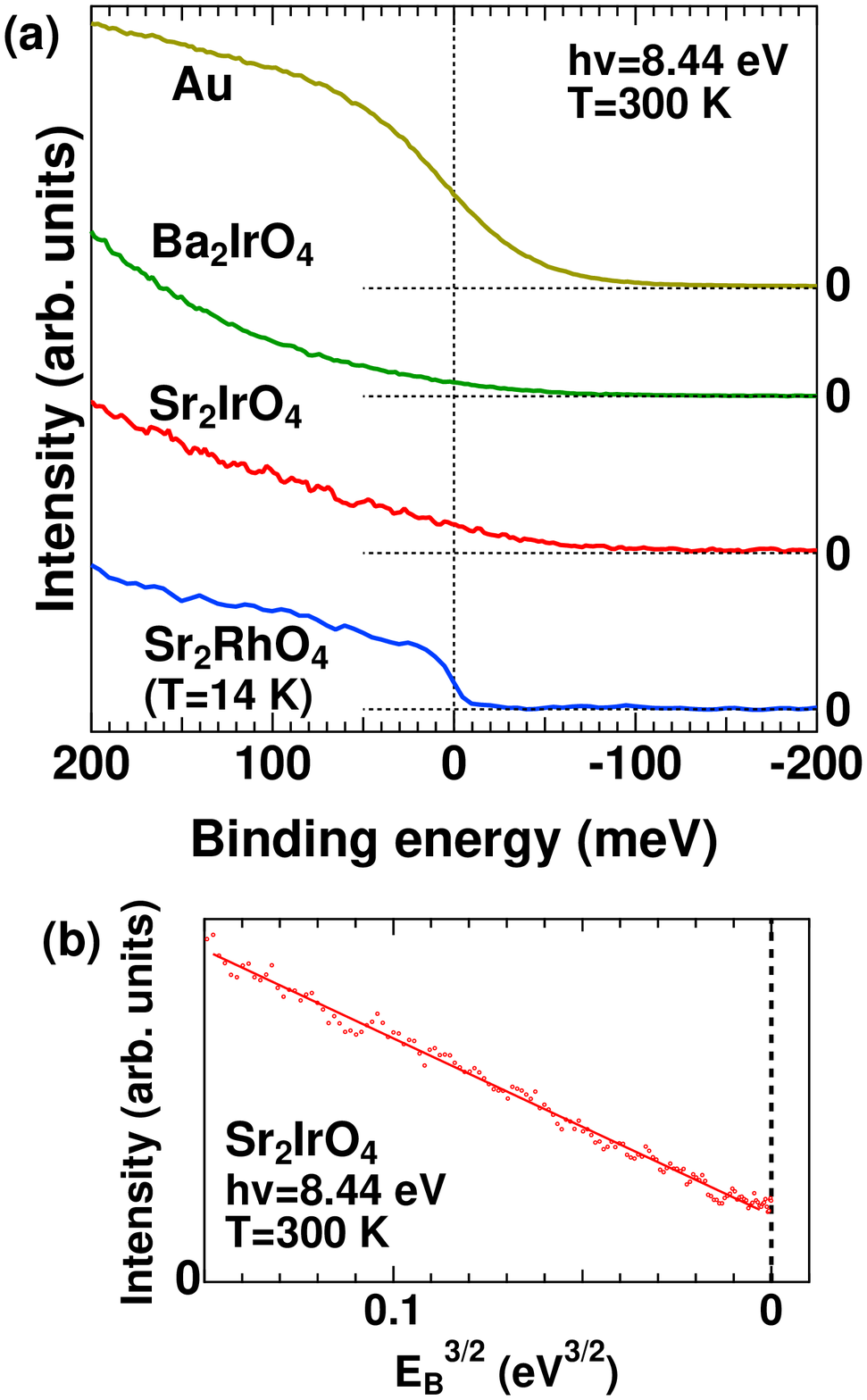}
\caption{
(Color online) 
(a) ELEPES spectra of  Sr$_2$IrO$_4$ and Ba$_2$IrO$_4$ near  $E_F$  at $T$=300~K ({\it above} $T_N$).
The ELEPES spectrum of {\it metallic} Sr$_2$RhO$_4$  at $T$=14~K is also shown as a reference.
(b) The ELEPES spectrum of Sr$_2$IrO$_4$ divided by Femi-Dirac distribution function.
The spectrum is plotted as a function of (binding energy)$^{3/2}$+ constant.
}
\label{Fig_3}
\end{figure}

Figure~\ref{Fig_1}(a) shows  valence-band PES spectra of Sr$_2$IrO$_4$ at three different $h\nu$s.
Both  Ir $5d$ and O $2p$ states have large spectral weight in the SXPES spectrum measured at $h\nu$=450~eV since the $\sigma$s of  these states are comparable  ($\sigma_{{\rm Ir}5d^5}/\sigma_{({\rm O}2p^6)_4}\simeq0.6$) and larger than those of other states.
In contrast, HAXPES spectrum measured at h$\nu\simeq8000$~eV originates mainly from the Ir $5d$  states due to the much larger $\sigma$ than those of other states, for instance, $\sigma_{{\rm Ir}5d^5}/\sigma_{({\rm O}2p^6)_4}\simeq80$.
Therefore, by comparing these  spectra,  one can conclude that  a large peak at the binding energy ($E_B$) of 3.4~eV(denoted as A), and two small peaks located at 10.3~eV(B) and 12.5~eV(C) contain strong O $2p$ components.
They are derived from the O $2p$ non-bonding states (A) and  Ir $5d$ $e_g-$ O $2p$ bonding states (B and C), respectively.
The HAXPES spectrum has  a two-peak structure in  which the lower- and higher-$E_B$ peaks (at $E_B\simeq$ 1~eV and 6~eV) originate from the Ir 5$d$ $j_{\it eff}-$ O $2p$ antibonding states and  bonding states, respectively.

We note that the photoelectron emission utilizing  linearly polarized light has angular distributions depending on the orbital symmetry of  electrons.
In the HAXPES experiments for Sr$_2$IrO$_4$ two different configurations with  $p$ and $s$ polarizations (the degrees of the linear polarization $P_L$s are +0.98 and -0.8, \cite{comment1} respectively) were employed as illustrated in Fig.~\ref{Fig_1}(d).~\cite{Sekiyama10}
One can expect to suppress the spectral weight derived mainly from  $s$ states in the spectrum measured with the $s$-polarized light.
Meanwhile,  the value of $\sigma_{{\rm Ir}5d^5}/\sigma_{({\rm O}2p^6)_4}$ for the $s$-polarized light is smaller ($\sim$20 at $h\nu=$8~keV in BL19LXU) than that for the $p$-polarized light ($\sim80$ at $h\nu=$8~keV in BL19LXU and $\sim90$ at $h\nu=$6~keV in P09).~\cite{Trzhaskovskaya}
As can be seen in Fig.~\ref{Fig_1}(c), the reduction of the spectral weight associated with the polarization change ($p\to s$)  has never been observed in the $j_{\it eff}$ anti-bonding bands ($0\leq E_B\leq\sim3$~eV), clearly indicating that the contribution of any $s$ states are negligibly small in this $E_B$ region.
Hereafter, we discuss the bulk Ir $5d$ $j_{\it eff}$ electronic structures based on the results of the HAXPES with the $p$-polarized light because of the  highest $\sigma$ ratio of Ir $5d$  to O $2p$ states.

Now one can directly compare the HAXPES spectra representing genuine bulk $j_{\it eff}$ states with the results of the LDA+DMFT calculations for  iridates.
Details of the calculation method have been reported elsewhere.~\cite{Arita12}
Figure~\ref{Fig_2}(a) shows the HAXPES spectra  in the $j_{\it eff}$ states. 
The LDA+DMFT spectra calculated for the antiferromagnetic insulating (AFI) ground states ($U$=1.96~eV for Sr$_2$IrO$_4$ and 1.6~eV for Ba$_2$IrO$_4$) and the paramagnetic insulating (PI) states ($U$=2.4~eV for Sr$_2$IrO$_4$ and 1.8~eV for Ba$_2$IrO$_4$) are also shown in Fig.~\ref{Fig_2}(a).
The calculated spectra of Sr$_2$IrO$_4$ and Ba$_2$IrO$_4$ for the PI phase (indicated by dotted lines, with larger $U$ values than the critical value $U_c$ for  Mott transition) can not reproduce the experimental spectra at all in the vicinity of $E_F$.
Therefore, the insulating nature of both iridates can never be described by the simple Mott picture.
Meanwhile, the calculated spectra for the AFI phase of Sr$_2$IrO$_4$ and Ba$_2$IrO$_4$ qualitatively reproduce  the experimental spectra,  supporting the ``Slater-dominant'' transition scenario for these compounds.~\cite{Arita12,comment3}

We note that the DMFT study strongly overestimated  their $T_N$s (810~K for Sr$_2$IrO$_4$ and  690~K for Ba$_2$IrO$_4$) without thought of both intersite fluctuations and interactions between crystal layers.
While the HAXPES spectra were measured at $T$=100~K (140-150~K below the real $T_N$ and this difference is defined as $\Delta T_{\rm PES}$), the LDA+DMFT spectra of Sr$_2$IrO$_4$ and Ba$_2$IrO$_4$ in Fig.~\ref{Fig_2}(a) were calculated for $T$=580~K (230 and 110~K below the calculated $T_N$s, respectively,  and these differences are defined as $\Delta T_{\rm DMFT}$).
The larger discrepancy between $\Delta T_{\rm PES}$ and $\Delta T_{\rm DMFT}$  in Sr$_2$IrO$_4$  yields less agreement between the experimental and theoretical spectra than in Ba$_2$IrO$_4$ in the empirical point of view.

The Slater picture  provides a metallic nature above $T_N$ unlike the Mott picture.
In general,  the  gap opens far above $T_N$ in Mott systems.~\cite{comment2}
In order to investigate the spectral behavior near $E_F$, the ELEPES spectra have been measured at $T$=300~K for  Sr$_2$IrO$_4$ and Ba$_2$IrO$_4$.
According to the ``universal'' curve  between   $\lambda_{\rm mp}$ in solids and the photoelectron kinetic energy,~\cite{Shirley}  photoelectrons excited by  Xe~I$\alpha$ light  are able to provide the information on  the bulk far from the surface.~\cite{Kiss05,Koralek06,Sato07}
The bulk sensitivity in  ELEPES is, however, known to depend on  individual  material systems, or in other words, on the dielectric function, band structures, and partial density of states (PDOS) near $E_F$.~\cite{Penn87,Offi08,Suga_book}
The low PDOSs near $E_F$ is thought to realize the enhanced $\lambda_{\rm mp}$ at $h\nu$=8.4~eV (Xe I$\alpha$).~\cite{Suga_book}
Figure~\ref{Fig_3} (a) shows the ELEPES spectrum of Sr$_2$RhO$_4$ which has a clear Fermi cutoff.
This indicates that  ELEPES has a longer probing depth than  low-$h\nu$ SXPES, in which the Fermi cutoff is unclear,  in this system and can reveal the bulk electronic structure.
It would be reasonable that one can  expect  higher bulk sensitivity for  Sr$_2$IrO$_4$ because of less conduction electrons.~\cite{Suga_book}

Figure~\ref{Fig_3} (a) shows the ELEPES spectra of Sr$_2$IrO$_4$ and Ba$_2$IrO$_4$ at $T$=300~K, that is, above $T_N$.~\cite{comment4}
Although the $\sigma$ of  Ir $5d$ states is much smaller than that of O $2p$ states ($\sigma_{{\rm Ir}5d^5}/\sigma_{({\rm O}2p^6)_4}\simeq0.06$) at $h\nu$=8.4~eV, the spectral weight near $E_F$ is  visible.
Significantly weak intensities at $E_F$ and no Fermi cutoff were observed unlike the gold, 
suggesting  ``bad metallic'' states in the paramagnetic phase.
This is consistent with the results of IPES at $T$=300~K, in which the spectra have  finite intensities at $E_F$ (see Fig.~\ref{Fig_2}(b)).
The  spectra of both iridates have some spectral weight  in the optical ``Mott'' gap of  200~meV below $E_F$.~\cite{Moon09,Kim12}
According to the LDA+DMFT results, the energy gap is closed above $T_N$ due to the strongly renormalized quasi-particle spectral weight at $E_F$. 

The ELEPES spectra of Sr$_2$IrO$_4$ follow  a 3/2 power law in the range from  $E_F$  to  $E_B\simeq$ 0.3~eV as shown in Fig.~\ref{Fig_3}(b).
The 3/2 power-law behavior was also  observed in other iridates, such as 
ferromagnetic BaIrO$_3$ and weak-ferromagnetic Y$_2$Ir$_2$O$_7$,
which was explained as a result of  a dominant contribution of ferromagnetic magnons to the electron excitation.~\cite{Maiti05,Singh08}
A recent resonant inelastic x-ray scattering study revealed  AF linear dispersion of magnons in Sr$_2$IrO$_4$,~\cite{Kim12} indicating that one could observe some features derived from the AF magnon excitation rather than  the ferromagnetic one.
In contrast, the spectra of Ba$_2$IrO$_4$ have never followed the power low (not shown here).
It is inferred that the spectral shape in  ELEPES is affected by the unoccupied density of states and  matrix element effects as well as electron correlations in the final state
 unlike in the high-energy PES.
Although the origin of the different behaviors between these iridates remains to be clarified,
it  is probably associated with the unoccupied Sr $4d$ and Ba $5d$ bands (at $\sim8$~eV above $E_F$), crystal symmetries, and the rotation of  IrO$_6$ octahedrons.


In conclusion, the HAXPES  experiments have revealed the bulk insulating nature of the layered perovskite iridates  Sr$_2$IrO$_4$ and Ba$_2$IrO$_4$.
The simple Mott picture cannot reproduce the overall valence band spectral features.
ELEPES demonstrated   both Sr$_2$IrO$_4$ and Ba$_2$IrO$_4$ have   ``bad metallic'' states in the paramagnetic phase,
suggesting   strongly renormalized metallic states  above  the N\'eel temperature as predicted by the LDA+DMFT calculations.


\ \\
\ \\
\ \\
\ \\
\ \
We would like to thank  T.~Aso, Y.~Nishitani, T.~Mori, and Y.~Matsui in Konan University, S.~Kitayama and T.~Matsumoto in Osaka University, J.~Kodama  and H.~Nagata in Hiroshima University for supporting experiments, R.~Arita and M.~Imada in University of Tokyo for providing the results of band structure calculations and fruitful discussion, and N.~Tomita in Yamagata University for fruitful discussion.
The soft x-ray experiments  at SPring-8 and the IPES experiments at HiSOR were performed  with the approval of the Japan Synchrotron Radiation Research Institute (JASRI)
(Proposal No.~2010A1227) and  the approval of Hiroshima Synchrotron Radiation Center (Proposal No.~12-B-29), respectively, 
under the support of Grant-in-Aid for Scientific Research for Young Scientists (B) (No.~23740244)
from the Ministry of Education, Culture, Sports, Science, and Technology, Japan, and Research Foundation for
Research Institute of Konan University.
Work in W\"{u}rzburg is supported by DFG (FOR 1162). 
The HAXPES instrument at beamline P09 is jointly operated by
the University of W\"{u}rzburg (R. Claessen), the University of Mainz (C. Felser), and DESY. 
Funding by the Federal Ministry of Education and Research (BMBF) under Contracts
No.~05KS7UM1, No.~05K10UMA, No.~05KS7WW3, and No.~05K10WW1 is gratefully acknowledged.





\begin{references}
\bibitem{comment1}
$P_L$ is defined as ($I_p$-$I_s$)/($I_p$+$I_s$), where $I_p$ and $I_s$ stand for
the intensities  of $p$- and $s$-polarization components in the x-ray, respectively.
\bibitem{comment3}
The peak at $E_B$=2.8~eV in Ba$_2$IrO$_4$ originates from  O $2p$ non-bonding states and does  not appear in the calculated spectrum.
\bibitem{comment2}
For instance, V$_2$O$_3$ is a typical Mott system. But the energy gap opens suddenly at $T_N$, since it is identical to the MIT temperature  in this material.
Across the $T_N$ the photoemission spectral shape drastically changes all over the valence band.~\cite{Fujiwara11}
This is not the present case since we have never observed such sharp changes.
\bibitem{comment4}
At low temperature one can hardly remove the influence of the electrical charging effect caused by the photoelectron emission 
from the ELEPES spectra of the insulating iridates since the photoelectrons with very low kinetic energy are very sensitive to the effect.
\end{references}
\end{document}